\documentclass{article} 
\usepackage[utf8]{inputenc}
\usepackage{lineno,hyperref}

\modulolinenumbers[5]

\usepackage{cite}
\usepackage{appendix}
\usepackage{amsmath,amssymb}
\usepackage{graphicx}
\usepackage{dcolumn}
\usepackage{booktabs}
\usepackage{float}
\usepackage{multirow}
\usepackage{bm}
\usepackage{braket}
\usepackage{soul}
\usepackage{stackengine}
\usepackage[normalem]{ulem}

\usepackage{xcolor}     

\begin{document}

\begin{center}
{\Large \bf Mixed scalarization of charged black holes: from spontaneous to non-linear scalarization}
\vspace{0.8cm}
\\
{Zakaria Belkhadria$^{1,2,3,4}$ and
Alexandre M. Pombo$^5$}
\\
\vspace{0.3cm}
$^1${\small Département de Physique Théorique, Université de Genève, 24 quai Ernest Ansermet, CH-1211 Geneva 4, Switzerland}\\

$^2${\small Gravitational Wave Science Center (GWSC), Université de Genève, CH-1211 Geneva, Switzerland}\\

$^3${\small Dipartimento di Matematica, Università di Cagliari, via Ospedale 72, 09124 Cagliari, Italy}\\

$^4${\small INFN, Sezione di Cagliari, Cittadella Universitaria, 09042 Monserrato, Italy}

$^5${\small  CEICO, Institute of Physics of the Czech Academy of Sciences, Na Slovance 2, 182 21 Praha 8, Czechia}\\

\end{center}

\date{\today}

\begin{abstract}
     Scalarized black holes (BH) have been shown to form dynamically in extended-scalar-tensor theories, either through spontaneous scalarization -- when the BH is unstable against linear perturbations -- or through a non-linear scalarization. In the latter, linearly stable BHs can ignite scalarization when sufficiently perturbed. These phenomena are, however, not incompatible and mixed scalarization is also possible. We explore two aspects of the Einstein-Maxwell-Scalar model: solutions containing, simultaneously, linear (\textit{aka} standard) and non-linear scalarization; and the effects of having one of the coupling constants with an 'opposite sign' to the one leading to scalarization. Both points are addressed by constructing and examining the mixed scalarization's domain of existence. An overall dominance of the spontaneous scalarization over the non-linear scalarization is observed. Thermodynamically, an entropical preference for mixed over the standard scalarization (spontaneous or non-linear) exists. In the presence of counter scalarization, a quench of the scalarization occurs, mimicking the effect of a scalar particle's mass/positive self-interaction term. 
\end{abstract}

%
\section{Introduction}\label{S1}
%

With the recent observational data from the LIGO-VIRGO collaboration (\textit{e.g.}~\cite{abbott2016observation,abbott2021gwtc}) and the direct imaging by the Event Horizon Telescope~\cite{ball2019first,akiyama2022first}, a heightened interest in alternative black hole (BH) solutions to the standard general relativistic (GR) ones has emerged. Of particular interest are hairy black hole solutions (see ~\protect\cite{herdeiro2015asymptotically,volkov2018hairy} for a review). More specifically, we focus on black holes resulting from scalarization processes, often referred to as scalarized black holes (For a review of scalarization, see \cite{doneva2022scalarization}).

These emerge in scalar-tensor theories, as originally proposed by Damour and Esposito-Farèse~\protect\cite{damour1993nonperturbative} for neutron stars. Extension of this framework to extended Scalar-Tensor (eST) theories include models that allow spontaneous scalarization of black holes \cite{silva2018spontaneous,doneva2018new,antoniou2018evasion}. In such models, the scalar field is non-minimally coupled to the model's invariants  and scalar perturbations of the vacuum black hole solutions can ignite the growth of scalar hair around the BH. The resulting scalarized black holes may present significant deviations compared to standard vacuum General Relativity (GR) solutions, offering deeper insights into the nature of gravity and particle physics.

    A family of eST theories that have undergone extensive analysis is the  Einstein-Maxwell-Scalar (EMS) model \cite{myung2019instability,herdeiro2018spontaneous,konoplya2019analytical,gan2021photon,gan2021photon2,guo2021scalarized,peng2020scalarization,myung2021scalarized,khalil2019theory,zhang2022critical,guo2022thermodynamics,hod2019spontaneous,hod2020analytic,hod2022spin,blazquez2020critical,kiorpelidi2023scalarization}. In the latter, scalarization is triggered by a non-minimal coupling between a real scalar field, $\phi$, and the Maxwell invariant, $\mathcal{I}=F_{\mu \nu}  F^{\mu \nu}$ (with $F_{\mu \nu}=\partial_\mu A_\nu-\partial_\nu A_\mu$ is the Maxwell tensor), through a coupling function $f(\phi )$ 
	\begin{equation}\label{E1}
	 \mathcal{S}=\int d^4 x\sqrt{-g} \left[ R - 2\, \partial _\mu \phi \partial ^{\mu} \phi - f (\phi)\mathcal{I} \right]\ ,
	\end{equation}
    and minimally coupled with the Ricci scalar, $R$, associated with the metric ansatz $g_{\mu\nu}$.  Scalarization is triggered by a sufficiently large charge-to-mass ratio. Although BHs within this model are considered of less astrophysical significance\footnote{In a dynamical astrophysical environment, the presence of plasmas around the BH leads to prompt discharge. Alternatively, the neutralization can occur through Hawking charge evaporation \cite{gibbons1975vacuum}.} compared to those in other eST models such as the model of Einstein Scalar-Gauss-Bonnet (EsGB) gravity \cite{doneva2018new,silva2018spontaneous,antoniou2018evasion}, their relative computational simplicity has proven valuable to developing a better insight into the scalarization phenomenon \cite{herdeiro2021spontaneous,zhang2021dynamical,zhang2022dynamical,yao2021scalarized,antoniou2023probing,jiang2023type,luo2022dynamical,niu2022dynamical,lai2022spin,xiong2022dynamical,chen2023time, promsiri2023scalarization, corelli2021challenging, chen2023nonlinear, xu2023tachyonic, guo2023scalarized}.
    
    In particular, the simplicity associated with the EMS model has allowed the study of several coupling functions~\cite{herdeiro2018spontaneous,fernandes2019spontaneous,astefanesei2019einstein}, from which two classes of solutions were found\footnote{The same classification also exists for Scalar-Gauss-Bonnet models~\cite{doneva2022beyond,pombo2023effects}}: class I (or dilatonic-type) and class II (\textit{aka} scalarization type). While in the former, \(\phi = 0\) is not a solution of the field equation\footnote{For the EMS model, there is an exceptional case: if $Q =P$, $\phi=0$ solves this class so that the dyonic, equal charges Reissner-Nordstrom BH is a solution.}, in the latter, the trivial scalar field \(\phi = 0\) is a solution and scalarized solutions can result from perturbations of the vacuum BH solution \footnote{A similar phenomenon was observed for a vector field instead of a scalar field \cite{oliveira2021spontaneous,barton2021spontaneously}, however, these seem to be prone to ghost instabilities~\cite{pizzuti2023spooky}}. This demands that

	\begin{equation}\label{E1.2.8}
		 \frac{df (\phi)}{d \phi}\bigg |_{\phi=0}= 0 \ .
	\end{equation}
    This condition is naturally implemented, for instance, if one requires the model to be $\mathbb{Z}_2$-invariant under $\phi\rightarrow -\phi$. Scalarized solutions can be further divided into two sub-classes. 

A necessary conditions for a tachyonic instability (and associated linear instability) to arise is
	\begin{equation}\label{Etaq}
	   \frac{d^2 f(\phi)}{d \phi^2}\bigg |_{\phi=0} \neq 0 \ ,
	\end{equation}
    and with the opposite sign of $\mathcal{I}$. Under these conditions the scalar hair around the BH grows spontaneously from a perturbed vacuum Reissner-Nordstrom BH (RN BH). In this case, the scalarized solutions bifurcate from the vacuum RN BH solutions. These are known as class II.A or spontaneous/normal scalarization.

A second family of scalarized solutions can also emerge even if BH are stable against linear perturbations
\begin{equation}\label{E3}
     \frac{d^2 f(\phi)}{d \phi^2}\bigg |_{\phi=0} = 0 \ .
\end{equation}
These are class II.B or non-linear scalarized solutions and may occur when the BH is sufficiently perturbed.

These two scalarization types, while distinct, can coexist simultaneously in what can been termed `mixed scalarization' . Such scenario has already been investigated in the context of (EsGB) theories \cite{doneva2022beyond,pombo2023effects}

    A possible coupling function compatible with both spontaneous and non-linear scalarization is
	\begin{equation}\label{mixedc}
	   f(\phi)=e^{-\alpha \phi ^2 -\beta \phi ^4}\ ,
	\end{equation}
	where $\alpha$ and $\beta$ are dimensionless coupling constants, with $\alpha <0$ and $\beta<0$. The sign of the coupling parameters $(\alpha,\ \beta)$ was chosen in accordance with the literature. Pure -- \textit{aka} non-mixed -- spontaneous (non-linear) scalarization being recovered when $\beta = 0$ ($\alpha =0$).

   The existence of scalarization is highly dependent on the sign of the spontaneous/non-linear function parameters ($\alpha,\,\beta$). While for pure scalarization the sign of either parameter is well defined, the presence of an additional scalarization mechanism, say spontaneous scalarization ($\alpha$), allows $\beta$ to have the ``wrong'' sign. We call this counter-scalarization.

    The objective of this work is twofold: first, to study the interplay between spontaneous and non-linear scalarization in the mixed scalarization scenario; second, to examine the effects of incorporating a counter-scalarization term into the coupling function. 
    
    The paper is organized as follows: Sec.~\ref{S2} introduces the basics of the EMS model, including a description of the equations of motion, boundary conditions, and relevant relations. Sec. \ref{S3} is dedicated to the numerical results. These include the computation of the domain of existence, Sec.~\ref{S3.2}, and both the entropical, Sec.~\ref{S3.2}, and perturbative, Sec.~\ref{S3.3}, stabilities. The paper concludes with some final remarks in Sec.~\ref{S4}.

    Throughout this paper, we set $16\pi G = 1=c$ for convenience. The spacetime signature is chosen to be $(-,+,+,+)$. We focus exclusively on spherically symmetric solutions, which implies that the metric and matter functions depend solely on the radial coordinate. For simplicity in notation, once a function is introduced with its radial dependency, such as $X(r)$, we will subsequently denote it by $X$ with the understanding that it is a function of $r$. Derivatives with respect to the radial coordinate $r$ and the scalar field $\phi$ are represented by $X' \equiv \frac{dX}{dr}$ and $X_{,\phi} \equiv \frac{dX}{d\phi}$, respectively.
%
\section{Black holes in the EMS model}\label{S2}
%
   As already stated in the Introduction (Sec.~\ref{S1}), in this work we are going to restrict ourselves to spherically symmetric EMS models described by action \eqref{E1}. For the line element, let us consider a standard metric ansatz that is compatible with spherical symmetry and has two unknown functions
    \begin{equation}\label{E2}
	 ds^2 = - N(r) e^{-2  \delta (r)} dt^2+\frac{dr^2}{N(r)}+r^2 \big(d\theta ^2 +\sin ^2 \theta\,  d \varphi ^2\big)\ ,~~{\rm with}~~N(r)\equiv 1-\frac{2\, m(r) }{r},
    \end{equation}
where $m(r)$ is the Misner-Sharp mass function~\cite{misner1964relativistic}, and $\delta (r)$ is an unknown metric function. For spherically symmetric and electrically charged BHs\footnote{While a magnetic charge would also be compatible with spherical symmetry, it is not considered in this context - see~\cite{astefanesei2019einstein} for magnetically charged BHs in this context.}, the electrostatic 4-vector potential, $A(r)=V(r)\,dt$, and the scalar field is solely radial-dependent $\phi(t,r,\theta,\varphi)\equiv\phi (r)$. The absence of angular dependence allows one to obtain the effective Lagrangian,

    \begin{equation}\label{Lag}
     \mathcal{L}_{\rm eff} =e^{-\delta }m'-\frac{1}{2}r^2 e^{-\delta}N\phi'^{2}+\frac{1}{2}f(\phi) e^{\delta }r^2 V'^{2 }\ .
    \end{equation}
    Variation of the effective Lagrangian with respect to the metric ($m,\ \sigma$) and matter functions ($V,\ \phi$) yields the field equations:
    \begin{equation}\label{ODE}
        \begin{aligned}
         & m'=\frac{r^2N \phi '^2}{2} +\frac{Q ^2}{2r^2 f(\phi)} \ , \qquad \delta'+r \phi '^2=0 \ , \qquad  V'  = -\frac{Q e^{-\delta}}{f(\phi) r^2} \ ,\\
         & \phi '' +\frac{1+N}{rN}\phi'-\frac{Q ^2}{r^3N f(\phi)}\left(\phi'-\frac{f_{, \phi}(\phi)}{2r  f(\phi)}\right)=0\ .
        \end{aligned}
    \end{equation}	
    Where the electrostatic potential $V$ is under a first integral, which was used to simplify the remaining equations. The constant of integration is interpreted as the electric charge, \( Q \).   

     To solve the set of four coupled ordinary differential equations \eqref{ODE}, one must implement the appropriate boundary conditions. At the horizon, $r=r_H$, the field equations can be approximated by a power series expansion in \( r - r_H \) as
    \begin{equation}\label{BC}
        \begin{aligned}
         & m = \frac{r_H}{2}+ \frac{Q ^2}{2r_H ^2 f(\phi_0)} (r-r_H) +\cdots\ \ , \qquad
        \delta   = \delta _0 -r_H\bigg(\frac{Q ^2}{2 r_H \big(Q ^2-r_H^2 f (\phi_0)\big)} \frac{ f_{,\phi} (\phi_0)}{ f (\phi _0)}\bigg)^2\, (r-r_H)+\cdots\ ,	\\
        & \phi =  \phi _0 + \frac{Q ^2 f_{,\phi} (\phi_0)}{2 r_H f (\phi _0) \big(Q ^2-r_H^2 f (\phi_0)\big)} (r-r_H)+\cdots\ , \qquad \qquad \ \ \ \ \ V = -\frac{e^{-\delta _0 }Q}{r_H ^2 f (\phi_0)} (r-r_H)+\cdots\ ,
        \end{aligned}
    \end{equation}
   in terms of the two essential parameters $\phi_0$ and $\delta_0$, where the subscript $_0$ denotes functions evaluated at the horizon $r_H$. At spatial infinity, asymptotic flatness is ensured by a power series expansion in $1/r$.

    \begin{equation}\label{INF}
        \begin{aligned}
	   & m(r) =M- \frac{Q^2+Q_s^2}{2r}+\cdots\ ,\qquad \qquad \qquad   
	\delta (r) \approx  \frac{Q_s^2}{2\, r^2}+\cdots\ ,\\
	   & \phi (r) = \frac{Q_s}{r}+\frac{MQ_s}{r^2}+\cdots\ ,\qquad \qquad \qquad \quad V(r) =\psi_e+\frac{Q}{r}+\cdots \ ,
        \end{aligned}
    \end{equation}
    with \( M \) representing the ADM mass, \( Q \) the BH's electric charge, and \( Q_s \) the scalar ``charge''\footnote{The term scalar ``charge'' is used due to the similar radial decay to a true electric charge, not because of an associated conserved Noether current.}, while \( \psi_e \) is the electrostatic potential at infinity. It is important to note that the so-called ``scalar'' hair associated with the EMS scalarization model is of a secondary nature and does not add any additional degree of freedom.

    Equation \eqref{ODE}, together with the boundary conditions arising from the power series expansion \eqref{BC} and \eqref{INF}, constitute a  Dirichlet boundary condition problem that must be numerically integrated (see Sec.~\ref{S3}).
    
%
    \subsection{Identities and physical quantities of interest}\label{S2.1}
%
    Scalarized solutions are physically characterised by the dimensionless quantities: charge-to-mass ratio, $q$, reduced horizon area, $a_H$, and reduced horizon temperature, $t_H$, 
        \begin{equation}
         q\equiv \frac{Q}{M} \ , \qquad  a_H\equiv \frac{A_H}{16\pi M ^2} = \frac{ r_H ^2}{4 M ^2}\ ,\qquad t_H\equiv 8\pi M T_H= 2MN'(r_H) e^{-\delta _0}\  \ ,
        \end{equation}
    where $A_H=4\pi r_H^2$ and $T_H=N'(r_H)e^{-\delta_0}/4\pi$ are the area and temperature of the BH's horizon, respectively. Regularity of the solutions is guaranteed by the Ricci scalar, $R$, and the Kretschmann scalar, $K\equiv R_{\mu \nu \delta \lambda} R^{\mu \nu \delta \lambda}$,

        \begin{equation}
            \begin{aligned}
             R &= \frac{N'}{r}\big( 3\, r\, \delta '-4\big) + \frac{2}{r^2}\Big\{ 1+N \Big[ r^2 \delta '' -\big( 1-r \delta '\big)^2\Big]\Big\} -N''\ , \\
            K &= \frac{4}{r^4}\big( 1-N\big)^2 + \frac{2}{r^2}\left[N'^{\, 2}+\big( N'-2N\delta'\big)^2\right] + \left[N''-3\,\delta' N' +2N\big( \delta'^{\, 2}-\delta''\big)\right]^2\ .
            \end{aligned}
       \end{equation}

    The accuracy of the numerically obtained solutions is guaranteed through the use of the so-called virial identity, Smarr law, and non-linear Smarr relation, which, besides their inherent physical significance, serve as crucial tests for the numerical accuracy and consistency of the solutions.
    
 The virial identity is obtained through a Derrick-like scaling argument~\cite{herdeiro2022deconstructing,oliveira2023convenient,herdeiro2021virial,derrick1964comments} and is given by
        \begin{equation}
         \int _{r_H} ^{\infty} dr \left\{ e^{-\delta }\, r^2\, \phi'^{\, 2} \left[ 1+\frac{2\, r_H}{r}\bigg(\frac{m}{r}-1\bigg)\right]\right\}
         = \int _{r_H} ^{\infty} dr \left[ e^{-\delta}\left(1-\frac{2\, r_H}{r}\right)\frac{1}{r^2}\frac{Q^2}{f(\phi)}\right]\ ,
        \end{equation}
    which is independent of the equations of motion and displays that scalarization can occur only in the presence of an electric charge \( Q \neq 0 \) in an EMS model. Since $1+\frac{2r_H}{r}\Big(\frac{m}{r}-1\Big) >0$, the left-hand side of the equation is strictly positive and can only be counterbalanced by a non-zero electric charge in the right-hand side.
    
   For this family of solutions, the Smarr law is found to be unaffected by the presence of the scalar hair, implying that the scalar field does not explicitly appear in the Smarr formula, \cite{herdeiro2018spontaneous}:
        \begin{equation}
         M = \frac{1}{2} T_H A_H + \psi_e Q\ .
        \end{equation}
    The first law of black hole thermodynamics for EMS black holes is expressed as
        \begin{equation}
         dM = \frac{1}{4} T_H \, dA_H + \psi_e \, dQ\ .
        \end{equation}
    At last, it can be shown that scalarized solutions also obey the so-called non-linear Smarr formula~\cite{herdeiro2018spontaneous},
        \begin{equation}
         M^2 + Q_s^2 = Q^2 + \frac{1}{4} A_H^2 T_H^2\ .
        \end{equation}

%
    \subsection{Coupling function}\label{S2.2}
%
    As previously mentioned, the choice of the coupling function is crucial in determining whether the RN BH is susceptible to scalarization, and hence one should be careful in designing the coupling function. Let us recap the conditions for scalarization. The first requirement is that the GR BH solution should also be a solution within the EMS model. Analysis of the field equations \eqref{ODE} shows that this can be secured by imposing \eqref{E1.2.8}. 
   This condition is naturally implemented if one requires the model to be $\mathbb{Z}_2-$invariant under the transformation $\phi\rightarrow -\phi$. The type of scalarization, on the other hand, is controlled by the second derivative of $f(\phi)$. For this, it is important to recall that the scalar field is described by the Klein-Gordon equation
      \begin{equation}\label{KG}
        \Box \phi = \frac{\mathcal{I}}{4} f_{,\phi} .
      \end{equation}
   
   Let us now consider a small-$\phi$ expansion of the coupling function
        \begin{equation}\label{E1.2.9}
	   f(\phi)=f(0)+\frac{1}{2}\frac{d^2 f}{d \phi^2}\bigg |_{\phi=0}\phi^2+\cdots \ ,
	\end{equation}
    the Klein-Gordon equation \eqref{KG} linearized for small-$\phi$ reads:
	\begin{equation}\label{KGphi}
	   \big(\Box-\mu_{\rm eff}^2\big)\phi =0\ , \qquad {\rm where} 
         \qquad 
         \mu_{\rm eff}^2= \mathcal{I}\, \frac{d^2 f(\phi)}{d \phi^2}\bigg |_{\phi=0}  \ .
	\end{equation}
	The instability arises if the scalar field's effective mass $\mu_{\rm eff}^2<0$ (\textit{aka} tachyonic mass), which, in particular, requires $f_{,\phi\phi}$ to obey \eqref{Etaq} and with the opposite sign of $\mathcal{I}$. This constitutes the set of solutions known as normal/spontaneous scalarization associated with class II.A, of which an exemplary function is
    \begin{equation}
        f(\phi) = e^{-\alpha \phi ^2}\ ,
    \end{equation}

    Class II.B on the other hand, occurs when vacuum solutions are linearly stable and no tachyonic instability exists, (see Eq.~\eqref{E3}). This condition is easily satisfied by considering higher order terms in the expansion of $f(\phi) $ such as    
    \begin{equation}
        f(\phi) = e^{-\beta \phi ^4}\ .
    \end{equation}
    An exemplary function that simultaneously exhibits tachyonic and non-linear instabilities is \eqref{mixedc}, $f (\phi) = e^{-\alpha \phi^2-\beta \phi^4}$. 
    In this function, the presence of a positive linear term ($\alpha > 0$) can counteract the usual scalarization process, as seen in the term $e^{-\alpha \phi^2}$ when $\alpha$ is positive. Conversely, a positive non-linear term ($\beta > 0$) in $e^{-\beta \phi^4}$ can similarly introduce opposing effects to scalarization.

It is feasible to have scalarized solutions where these counteracting terms, both the linear term $\alpha > 0$ and the non-linear term $\beta <0$, coexist and influence the overall scalarization dynamics. Such scenarios demonstrate the complex interplay between different scalarization mechanisms and their collective impact on the behavior of scalarized black holes.

Analysis of \eqref{KGphi} reveals that while these counteracting terms do not directly induce scalarization, they significantly modify the scalar field's dynamics: a positive linear term ($\alpha > 0$) functions as a mass term for the scalar field, and a positive quartic term ($\beta >0$) acts as a positive quartic self-interaction, both being proportional to the BH's electric charge due to the non-minimal coupling with the Maxwell invariant. Investigating their combined effect on scalarization phenomena is a pivotal aspect of this work.

%
\section{Numerical results}\label{S3}
%
    The set of four coupled ODEs~\eqref{ODE}, with the proper boundary conditions~\eqref{BC}-\eqref{INF}, are solved numerically as a two-point boundary value problem. The chosen routines automatically impose the proper boundary conditions through a shooting method on the two unknown parameters $\phi _0 $ and $\delta _0$, with the maximum integration error and boundary conditions automatically ensured to be less than $10^{-15}$.

\begin{figure}[ht]
    \centering
    \includegraphics[width=0.45\textwidth]{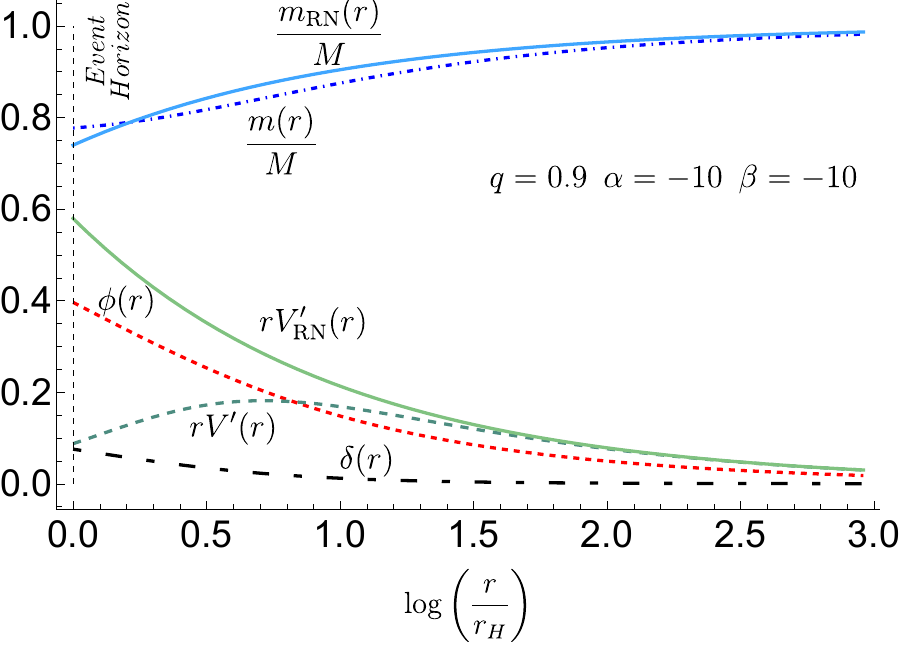}
    \caption{Solutions Profile (Radial functions) for a mixed scalarized black hole solution with parameters  \(\phi_0=0.40\), \(q=0.90\) and coupling constants \(\alpha=-10.0\) and \(\beta = -10.0\).}
    \label{F1}
\end{figure}

Physical accuracy of the solutions is further validated through the virial identity, with a relative error of $10^{-6}$, and the Smarr and non-linear Smarr relations, each with a relative error of $10^{-7}$. 

The resulting numerical solution's profile, depicted in Fig.~\ref{F1} for an exemplary solution characterized by the parameters \(\phi_0=0.40\), \(q=0.90\), \(\alpha=-10.0\), and \(\beta = -10.0\), shows that the scalar field $\phi(r)$ is monotonically decreasing with radius. At the horizon, $\phi_0$ is at its maximum, diminishing to zero at large radii. All defining radial functions, such as the metric function and the scalar field, converge to those of a comparable Reissner-Nordström (RN) black hole with similar global charges. Notably, the mass function approaches the ADM mass at infinity, confirming the consistency of these scalarized solutions.

All obtained solutions are regular at and outside the horizon, \(r_H\), with each scalarized black hole solution uniquely defined by the set of parameters \((r_H,\, \alpha,\, \beta,\, q)\).

%
    \subsection{Domain of existence}\label{S3.1}
First, let's outline the domain of existence for the two classes: Scalarized-Connected and Scalarized-Disconnected.
\textbf{Class II.A (Scalarized-Connected):} This class, depicted in Fig. 2 (left panel), illustrates scenarios of spontaneous scalarization occurring through non-minimal coupling of a scalar field to matter fields, typically via functions like \( e^{-\alpha \phi^2} \) where \( \beta = 0 \). Such scalarization gives rise to a domain where scalarized black holes coexist with Reissner-Nordström (RN) black holes. This domain is framed by an \textit{existence line}, comprising RN black holes that enable scalar field perturbations, and a \textit{critical line} that marks singular scalarized black hole configurations. Within this framework, particularly for \( q \leqslant 1 \), we observe a coexistence of scalarized and RN black holes sharing identical global charges. In these regions, scalarized black holes are generally entropically favored, hinting at their potential role as end-states in the evolution of linearly unstable RN black holes in the EMS model. Notably, along constant \( \alpha \) branches, the domain’s critical line reveals the possibility of scalarized black holes becoming overcharged as \( q \) extends beyond unity.\\
\textbf{Class II.B (Scalarized-Disconnected):} Conversely, illustrated in Fig 2 (right panel), this class covers nonlinear scalarization, arising even when background solutions are stable against linear perturbations. It requires significant nonlinear perturbations for scalar hair development and is associated with functions like \( e^{-\beta \phi^4} \) (with \( \alpha = 0 \)). The domain of existence for nonlinear scalarization is defined by different boundaries, emphasizing the influence of higher-order effects in the scalar field. This class features a novel two-branch structure of scalarized black holes, starting from the extremal RN black hole (the 'cold' branch) and extending into an over-extremal regime (the 'hot' branch). This unique structure leads to three different solutions for the same \( q \) in certain regions: two scalarized (cold and hot) and one standard RN black hole, a characteristic distinct from Class II.A.

\bigskip

\begin{figure}[ht]
     \centering
     \begin{minipage}{0.49\textwidth}
        \centering
    \includegraphics[width=\textwidth]{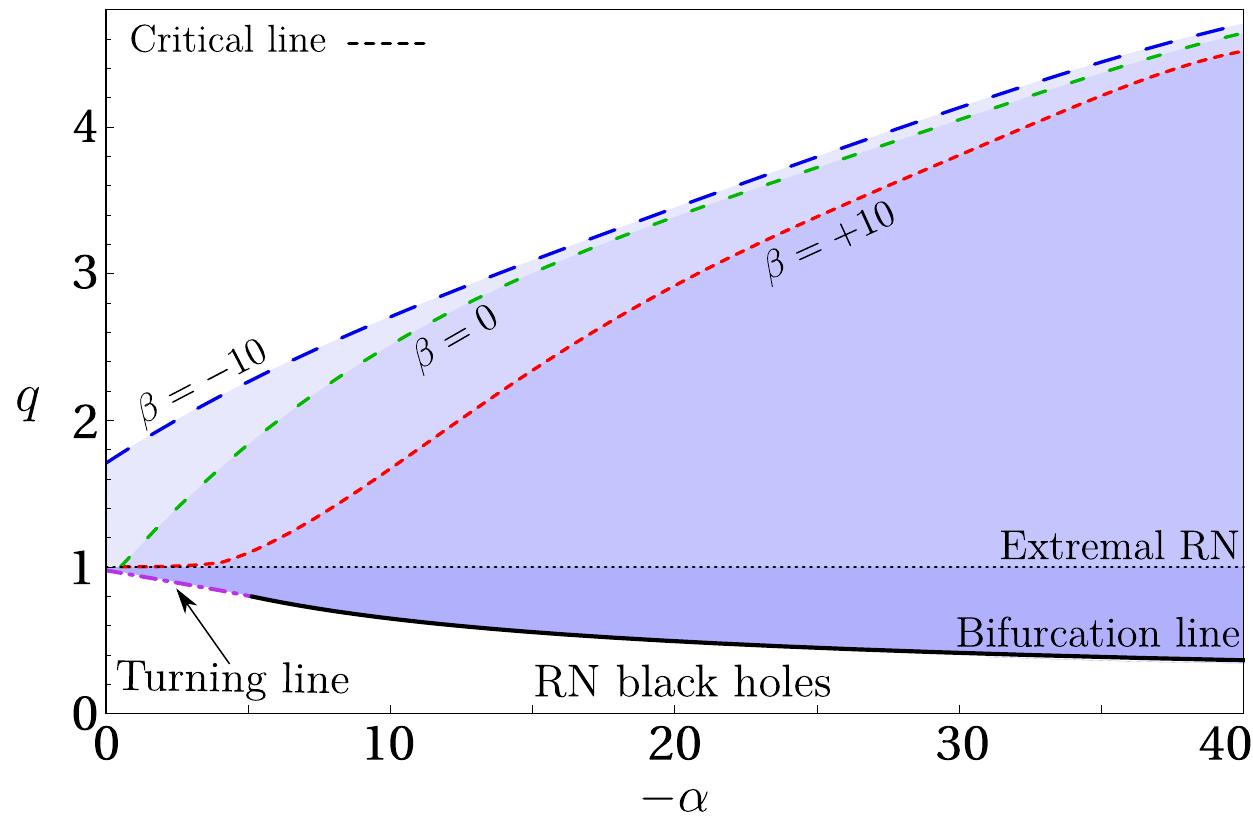}
    \end{minipage}\hfill
    \begin{minipage}{0.49\textwidth}
        \centering
\includegraphics[width=\textwidth]{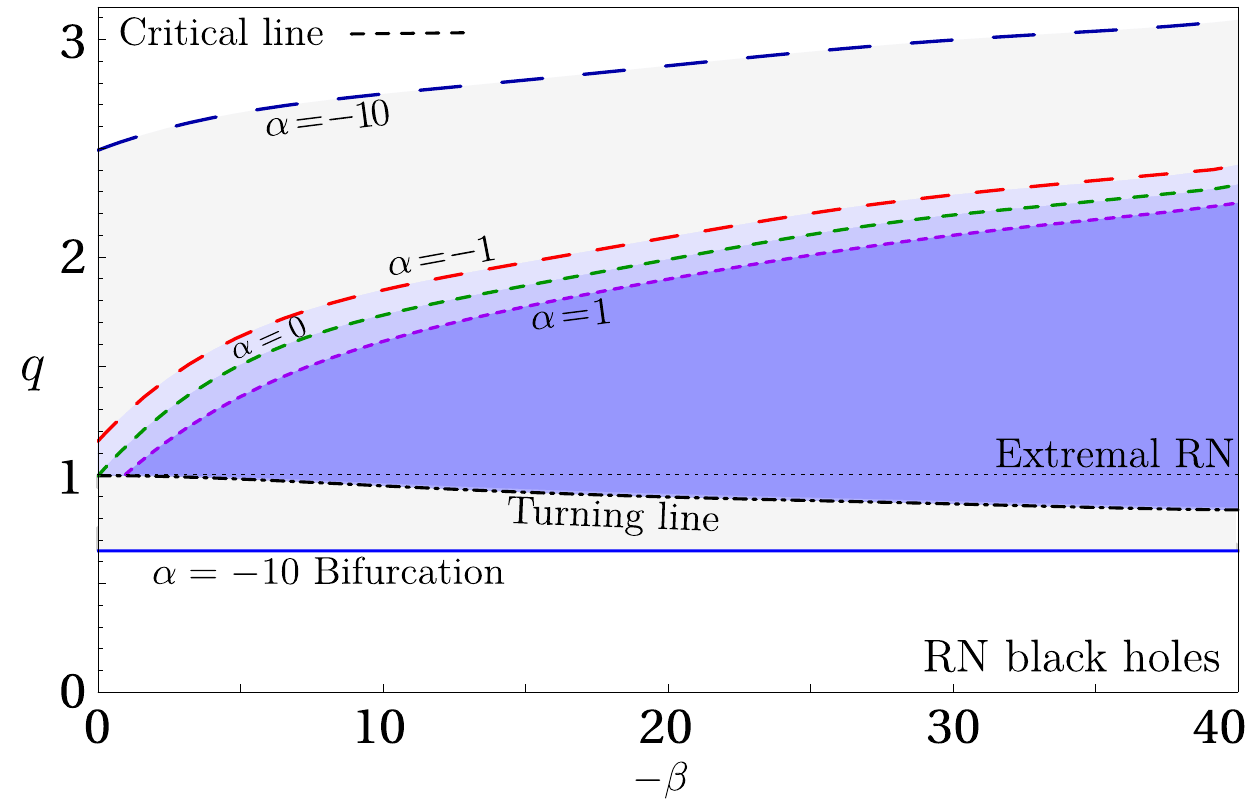}
    \end{minipage}
    \caption{Domain of existence of scalarized BHs in EMS models (shaded blue regions) with coupling function $f(\phi)=e^{-\alpha \phi^2-\beta \phi^4}$ as a function of $\alpha$ with $\beta =\{-10.0,\, 0,\, +10.0\}$ (left panel); or $\beta$ for $\alpha = \{-10.0,\,-1.0,\, 0,\, +1.0\}$ (right panel). The domain of existence is bounded from below by the existence/turning line (solid/dot-dashed line) and from above by the critical line (dashed line).}
    \label{F2}
\end{figure}

    Variation of the horizon radius, $r_H$, and coupling parameters, $(\alpha,\, \beta)$, for a fixed electric charge $Q$, form the $3-$dimensional domain of existence that characterizes the EMS model under analysis. To avoid the additional complexity associated with the $3-$dimensional domain of existence, let us fix one of the coupling parameters (say $\alpha$) and vary the other ($\beta$) -- see Fig.~\ref{F2}.
	
\begin{figure}[ht]
    \centering
    \includegraphics[width=0.50\textwidth]{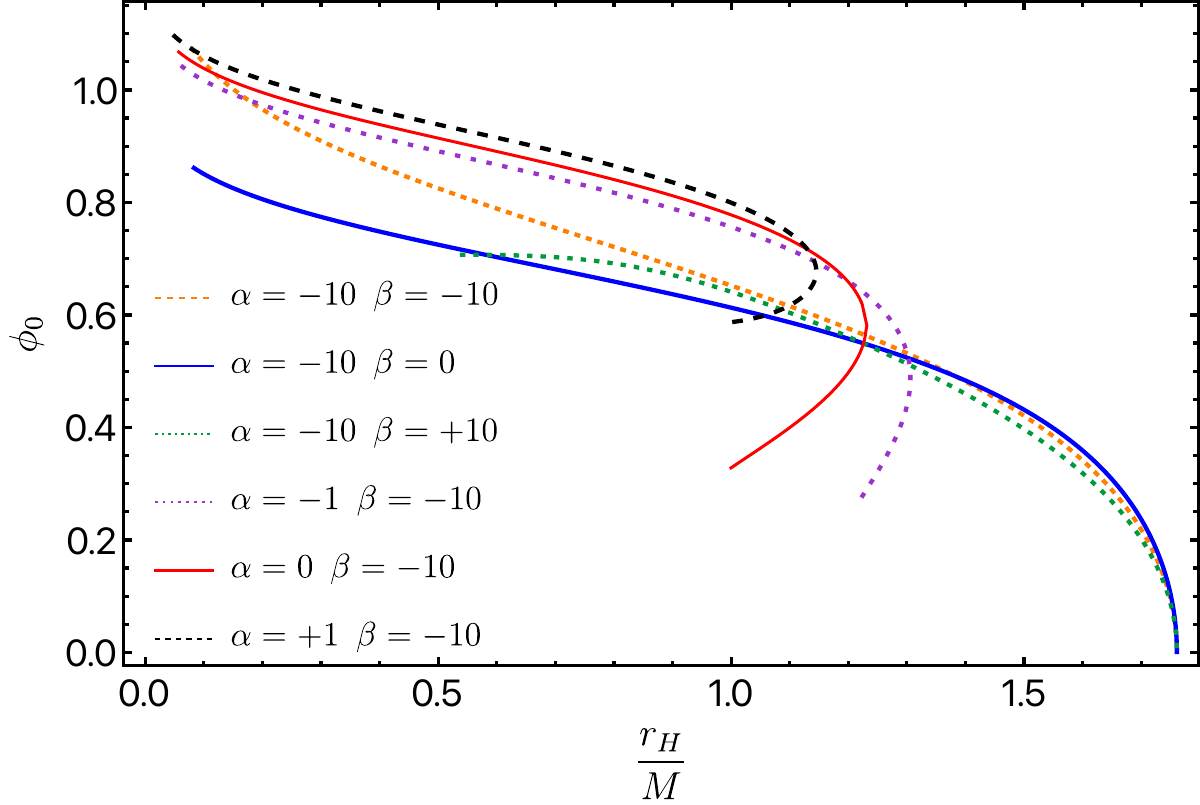}
    \caption{Scalar field horizon amplitude, \( \phi_0 \), as a function of the charge-to-mass ratio, \( q \), for a mixed scalarized BH for six different coupling parameters configurations: Orange dashed line for \((-10.0, -10.0)\), solid blue line for \((-10.0, 0.0)\), green dashed line with spacing for \((-10.0, +10.0)\), purple dashed line with closer spacing for \((-1.0, -10.0)\), solid red line for \((0.0, -10.0)\), and black dashed line for \((+1.0, -10.0)\).}
    \label{F3}
\end{figure}

	In Fig.~\ref{F2} is graphically represented the projection of the domain of existence of a scalarized BH as a function of $\alpha$ for three values of $\beta =\{-10.0,\, 0,\, +10.0\}$ (left panel); and as a function of $\beta$ for four values of $\alpha = \{-10.0,\,-1.0,\, 0,\, +1.0\}$ (right panel).
    A common feature of both domains of existence is a region with $q \leqslant 1$ -- where degeneracy occurs-- and a region with overcharged solutions $q>1$. The former region is limited from below by the bifurcation/turning line and from above by the extremal line $q=1$, and represents a region of the parameter space where, at least, two BH configurations with the same $q$ coexist: a bald RN BH and one or more scalarized BH solution. These solutions, however, possess different horizon radii (Fig.~\ref{F3}), temperatures and entropy (see Sec.~\ref{S3.2}). 

    The overcharged region goes from the extremal line, $q=1$, and is bounded from above by the critical line, which is highly coupling parameter's dependent: $q_{crit.} \equiv q_{crit.} (\alpha ,\beta)$. At the critical line, the solution's horizon radius tends to zero, $r_H \to 0$. In this region, no bald RN BH exist and no degeneracy between scalarized BHs was observed. After the critical line, no BH solutions exist.

    In addition, Fig.~\ref{F2} demonstrates a dominance of the tachyonic instability over the non-linear scalarization: for the same value of $\beta$ and $\alpha$, the domain of existence is bounded from below by the bifurcation line and has a $q_{crit.}$ comparable to the pure spontaneous scalarization, $\beta=0.0\,$. Such is expected since the maximum of the scalar field amplitude occurs at the horizon (see Fig.~\ref{F1}) and is smaller than unity $max(\phi)=\phi_0<1$, resulting in a decreased contribution of the higher powers of the scalar field\footnote{A similar behaviour was observed for the scalar-Gauss-Bonnet model \cite{doneva2022beyond}}.

    The lower bound, however, depends on the values of $\alpha$ and $\beta$. If $\beta \lesssim \alpha$, the scalarization is dominated by the tachyonic instability and the domain of existence is bounded from below by the bifurcation line; when $\beta \gg \alpha$, non-linear scalarization dominates and the domain of existence is bounded from below by the turning line (see Fig.~\ref{F2}). In this case, the tachyonic instability is not strong enough to sustain a negligible amount of scalar field (see Fig.~\ref{F3}), and a minimum value of $\phi_0$ exists. With the increase of $\alpha$, the size of the initial $\phi_0$ jump decreases until $\phi_0 \to 0$ and the tachyonic instability dominates.
	
    Let us now analyse the individual domains of existence. Starting with a fixed $\beta = \{-10.0, 0.0 ,+10.0\}$ and vary $\alpha$ -- see Fig.~\ref{F2} (left panel). In the majority of the domain of existence, solutions bifurcate from the RN BH at $q\leqslant 1$ -- where scalarized BHs with a negligibly amount of scalar field can exist -- and stops at the critical line, $r_H\to 0$. In between a continuously monotonic increase of the scalar field amplitude at the horizon and a reduction of the latter occurs. 

    As expected, the bifurcation line is insensitive to the non-linear term since the bifurcation is only dependent on the linear terms of $f_{,\phi}(\phi)$ that enter into the \textit{rhs} of the Klein-Gordon equation \eqref{KG}.

    The critical line, on the other hand, is highly dependent on the coupling parameters. The maximum charge-to-mass ratio for which the critical solutions exist, $q_{crit.}$, increases/decreases with the addition of a negative/positive $\beta$ term. While for $\beta<0$, a non-linear instability that amplifies the scalarization exists, a positive value has the opposite effect. The $\beta >0$ decreases the width of the domain of existence due to a decrease in $q_{crit.}$, resulting in a quench of the scalarization similar to the one observed for scalarization with a positive self-interaction~\cite{fernandes2020einstein}.
	
    Observe now the case with fixed $\alpha =\{-10.0,-1.0, 0.0,+1.0\}$ and varying $\beta$ -- Fig.~\ref{F2} (right panel). In this case, for all the chosen $\alpha\neq -10.0$, solutions start at the extremal line $q=1$ for which a minimal, non-negligible, amount of scalar field exists around the BH, $\phi_0 \neq 0$. \\
    By analyzing Fig.~\ref{F3}, we can notice that the BH's scalar field's associated jump high decreases (increases) with the addition of the $\alpha <0$ ($\alpha >0$) term 
    
    Following the initial jump in the scalar field amplitude at the horizon, solutions observe a simultaneous increase of $\phi_0$ and $r_H$ until a maximum mass, $r_H$ is reached -- Fig.~\ref{F3} $\beta=-10$ lines. After this point, a second branch with decreasing $r_H$ and increasing $\phi _0$ exists until the critical solution is achieved. The latter is known as the hot branch and is known to be stable for $\alpha =0$~\cite{blazquez2020einstein,blazquez2021quasinormal}; the former is known as the cold branch and is unstable (the denomination will be clear in the thermodynamics section \ref{S3.2}).

    As mentioned before, while the effect of a positive $\alpha$ is similar to scalarization by a massive scalar field, the positive $\beta$ mimics a positive self-interacting (attractive) potential. The present results follow the same pattern as the one presented for a massive/self-interacting scalar field in EMS \cite{fernandes2020einstein,zou2019scalarized} and scalar-Gauss-Bonnet models \cite{macedo2019self,doneva2019gauss,pombo2023effects}. In particular, a quench of the scalarization phenomena due to the decrease in the domain of existence width (see Fig.~\ref{F2}). However, due to the non-minimal coupling between the "mass" or "self-interaction" terms to the Maxwell invariant, the impact of the latter is proportional to the electric charge $Q$.

%
    \subsection{Thermodynamics}\label{S3.2}
\begin{figure}[H]
    \centering
    \includegraphics[width=0.49\textwidth]{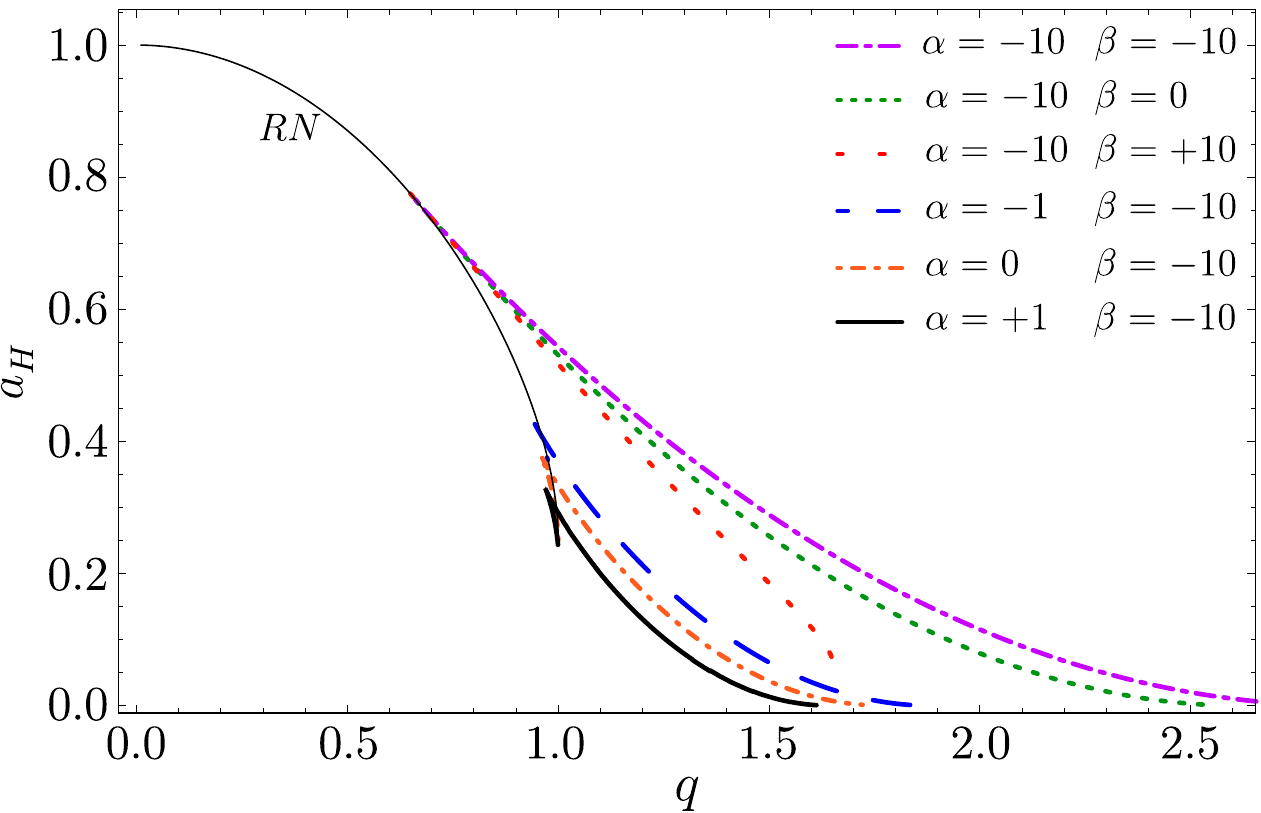}\hfill
    \includegraphics[width=0.49\textwidth]{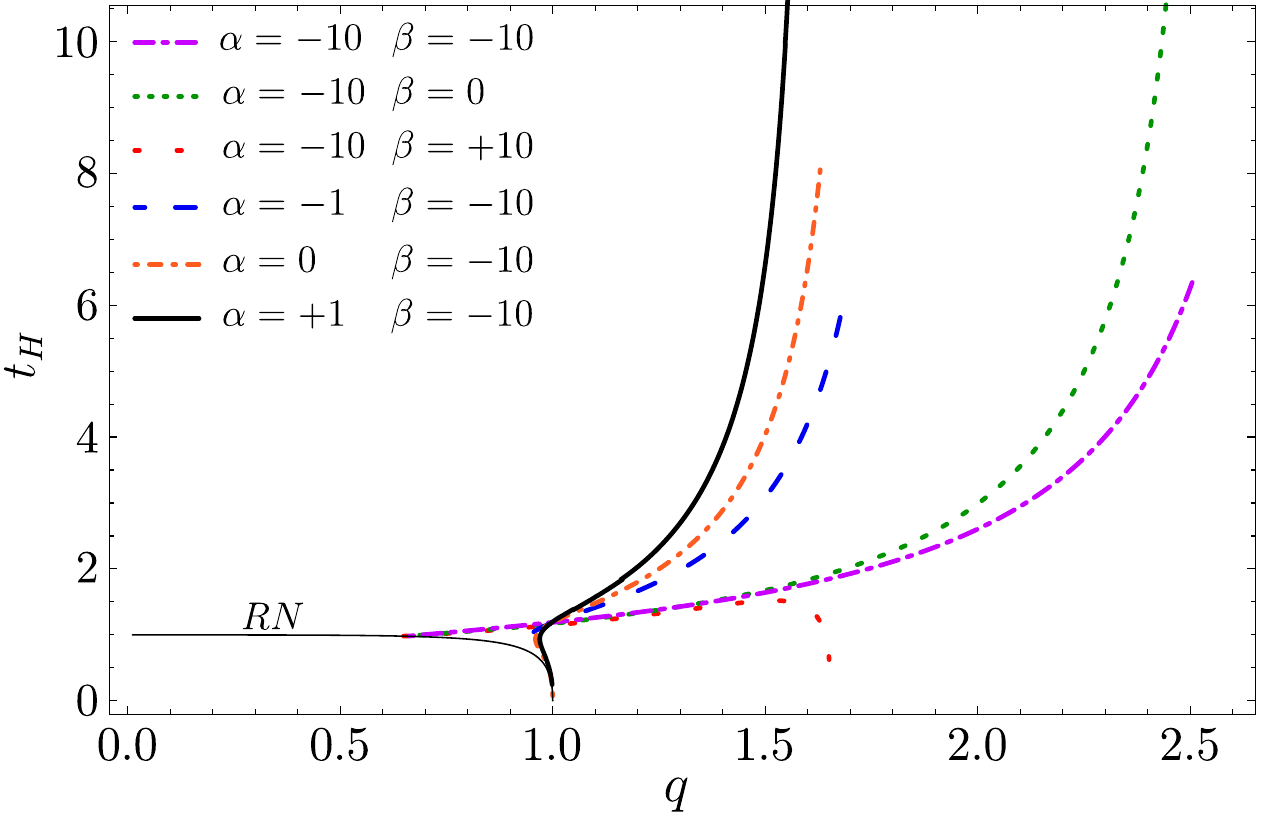}
    \caption{Reduced area, \( a_H \), (left panel) and reduced horizon temperature, \( t_H \), (right panel) as a function of the charge-to-mass ratio \( q \) for a mixed scalarized BH with six different coupling parameter's configurations \((\alpha, \beta)\): (dot-dashed purple) mixed scalarization \((-10.0, -10.0)\); (dotted green) pure spontaneous scalarization \((-10.0, 0.0)\); (dotted red) spontaneous scalarization \((-10.0, +10.0)\); (dashed blue) mixed scalarization with \((-1.0, -10.0)\); (dot-dashed orange) pure non-linear scalarization \((0.0, -10.0)\); mixed scalarization with \((+1.0, -10.0)\).}
    \label{F4}
\end{figure}

%
    A solution is said to be stable if it is simultaneously entropically preferable and stable against radial perturbations. The latter will be studied in Sec.~\ref{S3.3}. In EMS models, entropy is given by the Bekeinstein-Hawking formula~\cite{hawking1975particle,bekenstein2020black,majumdar1998black} and reduces to the analysis of the reduced horizon area, $a_H$, observe Fig.~\ref{F4} (left panel).
    Analysis of the horizon area shows that solutions dominated by spontaneous scalarization are always entropically preferable when compared with electro-vacuum GR; while non-linearly dominated solutions have a first branch that is everywhere entropically unfavourable (\textit{aka} cold branch), and a second branch which contains a set of solutions that are entropically preferable (\textit{aka} hot branch). The second branch's entropically non-preferable region decreases (increases) with the addition of $\alpha <0$ ($\alpha >0$). The terminology hot/cold comes from the second branch having a higher temperature than the first, see Fig.~\ref{F4} (right panel).
    
%
    \subsection{Comments on stability}\label{S3.3}
%
   
    Stability against radial perturbations is studied through a standard strategy that considers spherically symmetric, linear perturbations of the equilibrium solutions while keeping the metric ansatz \eqref{E2}, but allowing the functions $N,\, \delta,\, \phi,\, V$ time, $t$, dependent besides $r$:    
        \begin{equation}
         d s^{2}=-\tilde{N}(r, t)\, e^{-2\, \tilde{\delta}(r, t)} d t^{2}+\frac{d r^{2}}{\tilde{N}(r, t)}+r^{2}\left(d \theta^{2}+\sin ^{2} \theta\, d \varphi^{2}\right)    , \quad A=\tilde{V}(r, t)\, d t, \quad \phi=\tilde{\phi}(r, t)\ .
        \end{equation}
    Each function can be further expanded into the equilibrium solution (\textit{aka} ``bare'') that where obtained previously (Sec.~\ref{S3}-\ref{S3.2}), plus a perturbation term. The latter contains the time-dependence through a Fourier mode with frequency $\Omega$,
        \begin{equation}
            \begin{aligned}
             & \tilde{N}(r, t)=N(r)+\epsilon N_{p}(r) e^{-i \Omega t}\ , \quad \tilde{\delta}(r, t)=\delta(r)+\epsilon \delta_{p}(r) e^{-i \Omega t}\ , \\
             & \tilde{\phi}(r, t)=\phi(r)+\epsilon \phi_{p}(r) e^{-i \Omega t}\ , \quad \tilde{V}(r, t)=V(r)+\epsilon V_{p}(r) e^{-i \Omega t}\ ,
            \end{aligned}   
        \end{equation}
    where the subscript $_p$ denotes perturbations of the equilibrium solutions. The linearized field equations around the background solution yield the metric and $V_{p}(r)$ perturbations expressed in terms of the scalar field perturbation, $\phi _p (r)$,
        \begin{equation}
         N_{p}=-2\, r\, N \phi^{\prime}\, \phi_{p}\ , \qquad \delta_{p}=-2 \int d r\, r\, \phi^{\prime} \phi_{p}^{\prime}\ , \qquad V_{p}^{\prime}=-V^{\prime}\bigg[\delta_{p}+\phi_{p} \frac{f_{, \phi}(\phi)}{f(\phi)}\bigg]\ ,
        \end{equation}
    This leads to a single perturbation equation for $\phi_p (r)$ which can be written in a Schr\"odinger-like form by redefining $\Psi(r) = r\, \phi_{p}$ and inserting the 'tortoise' coordinate $x$ defined by $dx/dr = e^{\delta} / N$:
        \begin{equation}\label{E28}
         -\frac{d^{2}\Psi}{d x^{2}}+U_{\Omega} \Psi=\Omega^{2}\, \Psi\ ,
        \end{equation}

    with the perturbation potential $U_{\Omega}$ defined as:
        \begin{equation}
         U_{\Omega} \equiv \frac{e^{-2 \delta} N}{r^{2}}\left\{1-N-2\, r^{2} \phi^{\prime \, 2}-\frac{Q^{2}}{2 r^{2}}\left[\frac{2}{f(\phi)}\left(1-2\, r^{2} \phi^{\prime 2}\right)-\frac{2 f_{, \phi}^2(\phi)}{f^{3}(\phi)}+\frac{1}{f^{2}(\phi)}\Big(f_{, \phi \phi}(\phi)+4\, r\, \phi^{\prime} f_{, \phi}(\phi)\Big)\right]\right\}
        \end{equation}
    The potential $U_{\Omega}$, which is regular across the entire domain $-\infty < x < \infty$, diminishes to zero at both the black hole (BH) event horizon and infinity. A mode is classified as unstable if $\Omega^{2} < 0$, which, within the asymptotic boundary conditions of our framework, indicates a bound state. However, a standard quantum mechanical result dictates that equation \eqref{E28} will not exhibit bound states if $U_{\Omega}$ consistently exceeds its minimal asymptotic value, implying it must be positive (see \textit{e.g.} \cite{messiah1961quantum}). Therefore, a uniformly positive effective potential serves as evidence of mode stability against spherical perturbations.

 \begin{figure}[h]
         \centering
         \includegraphics[width=0.49\textwidth]{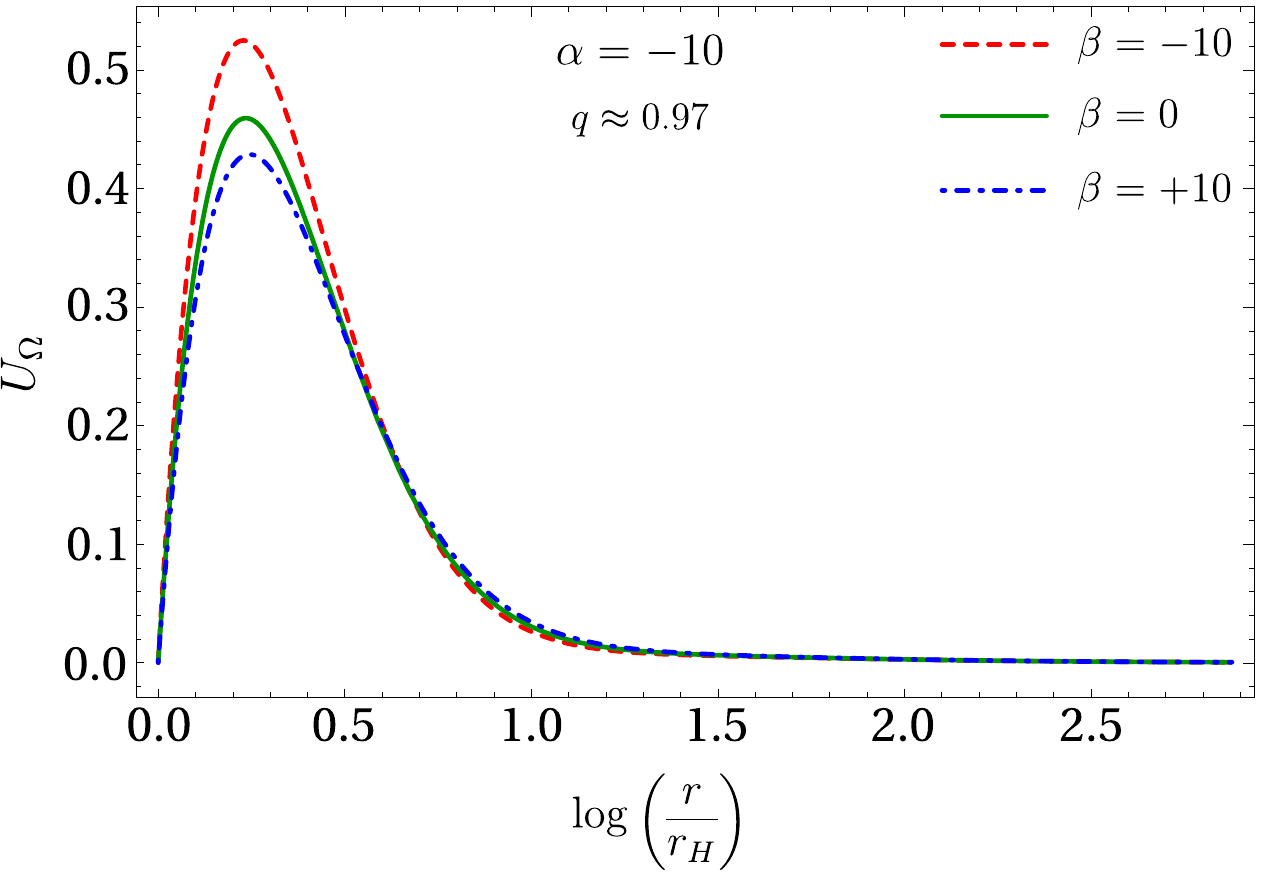}\hfill 
         \includegraphics[width=0.49\textwidth]{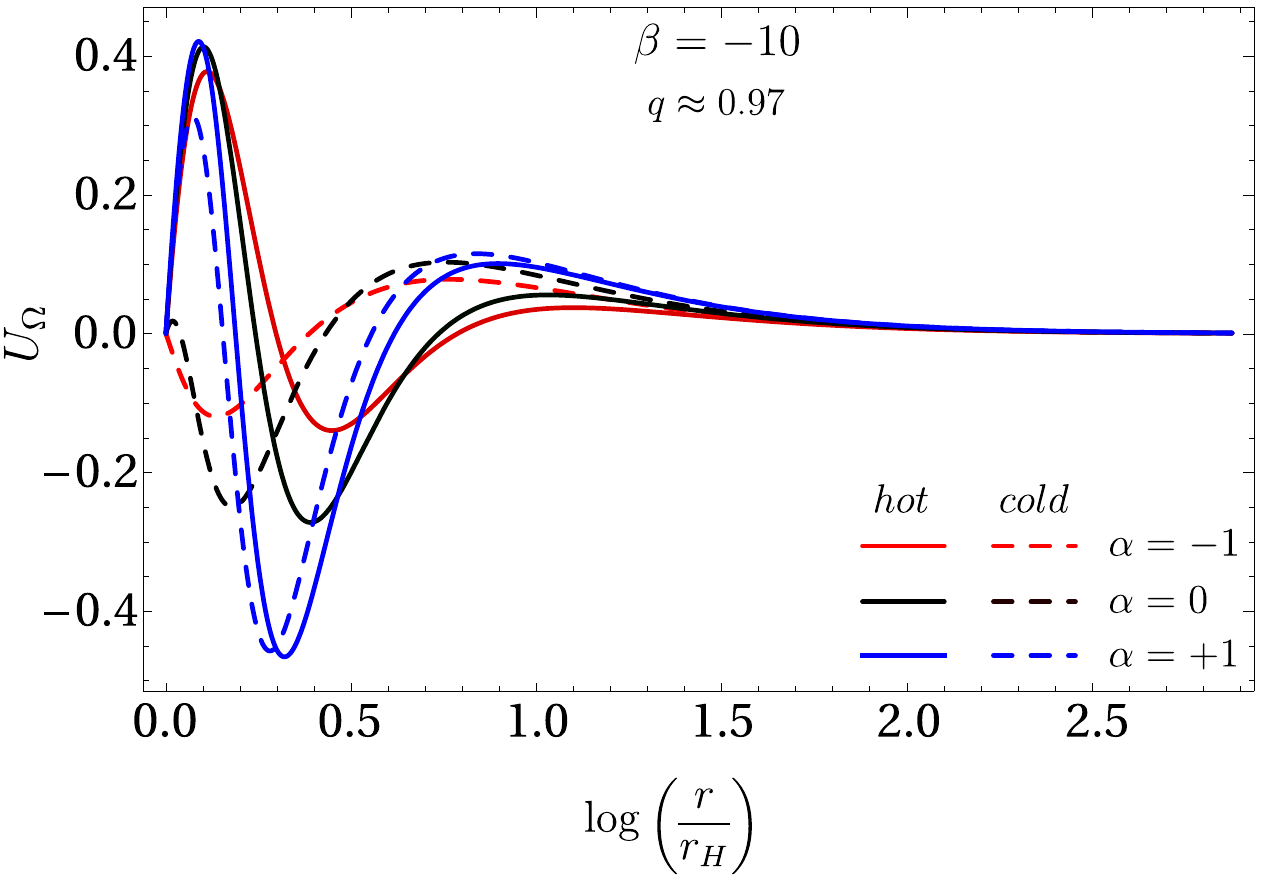}
         \caption{Effective potential for a set of scalarized RN BH solutions. (Left panel) Spontaneous scalarized dominated solutions with $\alpha=-10.0$ and (dashed red) $\beta=-10.0$ mixed scalarization, (solid green) $\beta=0.0$ pure spontaneous scalarization, and (dot-dashed blue) $\beta=+10.0$ counter-non-linear. (Right panel)  non-linear scalarization dominated solutions with $\beta=-10.0$ for both hot (solid) and cold (dashed) branches with (red) $\alpha=-1.0$ mixed scalarization, (black) $\alpha =0.0$ pure non-linear scalarization, and (blue) $\alpha =+1.0$ counter-spontaneous scalarization.}
         \label{F5}
        \end{figure}

    To discern the interplay between the parameters $\alpha$ and $\beta$ on the effective potential, we analyzed the profile of $U_{\Omega}$ for several scalarized solutions, holding $\alpha$ constant while varying $\beta$ -- Fig.~\ref{F5} (left panel)--, and vice-versa -- Fig.~\ref{F5} (right panel).   

    The effective potential for radial spherical perturbations of spontaneously scalarized dominated solutions corresponding to a fixed $ \alpha =-10.0$ -- Fig.~\ref{F5} (left panel) --, exhibit an everywhere positive effective potential, $U_{\Omega}>0$, suggesting the absence of instabilities. Additionally, one observes an increase (decrease) of the maximum value of $U_\Omega$ with the addition of the non-linear (counter-non-linear) parameter $\beta$.
    
    In the case of non-linearly dominated scalarized solutions with fixed $\beta =-10.0$ -- Fig.~\ref{F5} (right panel) --, for both the hot and cold branches, a region where $U_{\Omega}$ changes sign exists. Which, while not indicating instabilities, doesn't guarantee stability.
    
    The addition of a tachyonic term, $\alpha =-1.0$, to the non-linearly dominated scalarized solution reduces the amplitude of the negative $U_{\Omega}$ region, bringing it closer to the stable spontaneously lead scalarization, while a counter-scalarized term $\alpha =+1.0$ deepens it. 

    At last, it is important to note that a non-uniformly positive potential for the non-linear scalarized solutions is not a guarantee of instabilities. To further investigate these solutions, the application of more intricate methods like the S-deformation method~\cite{kimura2017simple,kimura2018robustness,kimura2019stability} is required. Such is beyond the scope of this paper (see \cite{blazquez2020einstein} for a similar study of the pure non-linear scalarization, and \cite{fernandes2020einstein,zou2019scalarized} for the impact of self-interacting/mass term).

\newpage
%
\section{Conclusions}\label{S4}
%

    In this work, we have investigated the interplay between the spontaneous and non-linear scalarization of charged black holes within the Einstein-Maxwell-Scalar model. The resulting mixed scalarized solutions possess both properties of pure non-linear and spontaneous scalarization.

	The interplay between the two types of scalarization shows a dominance of the spontaneus properties over the non-linear ones. In particular, the domain of existence for comparable coupling parameters possesses the same structure of the spontaneous scalarization with a slight increase in the width from the non-linear scalarization. The influence of non-linear scalarization becomes apparent only when its coupling parameter is significantly larger than that of spontaneous scalarization.

    The tachyonic instability associated with the spontaneous scalarization in the mixed coupling makes the black hole more susceptible to scalarization. As a result, it requires a weaker coupling of the scalar field to the Maxwell invariant to achieve the same level of scalarization.

	The presence of a mixed scalarization also allowed the study of a `counter-scalarization' term, wherein one of the scalarization parameters has the `wrong' sign and,

 instead of supporting/intensifying the scalarizaton, suppresses it. From the linearized KG equation, one observes that these terms possess the properties of a scalar field's mass (for the linear parameter $\alpha >0$) or of a positive self-interaction (for the non-linear $\beta >0$). The resulting quench associated with these parameters mimics the effect of a scalar field's mass/self-interaction observed in previous studies \cite{fernandes2020einstein,zou2019scalarized}. However, due to the coupling to the Maxwell invariant, the effect is not constant.

    Thermodynamically, mixed scalarized solutions with higher negative values of both $\alpha$ and $\beta$ are favourable. Solutions dominated by spontaneous scalarization are entropically preferable over their General Relativity counterparts, while non-linear dominated solutions show mixed thermodynamic behaviour, \textit{i.e.} an entropically favourable and unfavourable regions.

    Perturbative stability analysis against radial perturbations indicated stability for dominant spontaneously scalarized black holes. In contrast, for non-linear dominated scalarization, such a conclusion is not possible to be made. Further studies into these must be made.

	However, it seems like the addition of the spontaneous scalarization parameter to the non-linear scalarization makes the resulting mixed scalarization tend to a more stable configuration.

    Future research directions include applying the S-Deformation method for an in-depth analysis of the radial stability of non-linearly dominated solutions. A comprehensive study of quasi-normal modes in mixed scalarized solutions could provide further insights extending the work of \cite{myung2019quasinormal,blazquez2021quasinormal,myung2019stability,guo2022quasinormal}. Additionally, incorporating rotation into these mixed scalarization configurations presents an intriguing avenue for exploration. 
%
\section*{Acknowledgments}
%
We would like to express our sincere gratitude to Eugen Radu for reading and commenting on the manuscript and for the insightful discussions. Our thanks also go to Pedro G.S. Fernandes and Nuno M. Santos for their valuable discussions. Z.B extends special thanks to the Gravitation Group of Aveiro (CIDMA) and to Carlos Herdeiro for their hospitality during the initial phase of studying EMS models. Z.B also gratefully acknowledges the networking support provided by the COST Action CA18108. A. M. Pombo is supported by the Czech Grant Agency (GA\^CR) under grant number 21-16583M.

\newpage

%


  \bibliographystyle{ieeetr}
  \bibliography{references}

%

\end{document}